\documentclass{article}
\usepackage{graphicx}
\usepackage{hyperref}
\usepackage{lineno}
\usepackage[OT1]{fontenc}
\usepackage{amsthm,amsmath,amssymb}
\usepackage[numbers]{natbib}
\usepackage{graphicx}
\usepackage{subfig}
\usepackage{titlesec}
\usepackage{mathrsfs}
\usepackage{multirow}

\RequirePackage{enumitem} 
\RequirePackage{mathtools}
\usepackage[doublespacing]{setspace}
\usepackage{geometry}
\numberwithin{equation}{section}
\theoremstyle{plain}

\titlelabel{\thetitle.\quad}
\def\bibindent{1em}

\makeatletter
\renewcommand\@biblabel[1]{} 
\renewenvironment{thebibliography}[1]
{\section*{\refname}%
	\@mkboth{\MakeUppercase\refname}{\MakeUppercase\refname}%
	\list{\@biblabel{\@arabic\c@enumiv}}%
	{\settowidth\labelwidth{\@biblabel{}}%
		\leftmargin\labelwidth
		\advance\leftmargin15pt
		\advance\leftmargin\labelsep
		\setlength\itemindent{-10pt}
		\@openbib@code
		\usecounter{enumiv}%
		\let\p@enumiv\@empty
		\renewcommand\theenumiv{\@arabic\c@enumiv}}%
	\sloppy
	\clubpenalty4000
	\@clubpenalty \clubpenalty
	\widowpenalty4000%
	\sfcode`\.\@m}
{\def\@noitemerr
	{\@latex@warning{Empty `thebibliography' environment}}%
	\endlist}
\renewcommand\newblock{\hskip .11em\@plus.33em\@minus.07em}
\makeatother

\newenvironment{bottompar}{\par\vspace*{\fill}}{\clearpage}
\renewcommand{\refname}{REFERENCES}
\DeclareMathOperator*{\argmin}{argmin}

\begin{document}

\title{\textbf{A two-stage Fisher exact test for multi-arm studies with binary outcome variables}}
\author{\textbf{M. J. Grayling\textsuperscript{1}, A. P. Mander\textsuperscript{1}, J. M. S. Wason\textsuperscript{1,2}}\\
	\small 1. Hub for Trials Methodology Research, MRC Biostatistics Unit, Cambridge, UK, \\ \small 2. Institute of Health and Society, Newcastle University, Newcastle, UK.}
\date{}
\maketitle

\noindent \textbf{Running Head:}  A two-stage multi-arm Fisher exact test\\

\noindent \textbf{Abstract:} In small sample studies with binary outcome data, use of a normal approximation for hypothesis testing can lead to substantial inflation of the type-I error-rate. Consequently, exact statistical methods are necessitated, and accordingly, much research has been conducted to facilitate this. Recently, this has included methodology for the design of two-stage multi-arm studies utilising exact binomial tests. These designs were demonstrated to carry substantial efficiency advantages over a fixed sample design, but generally suffered from strong conservatism. An alternative classical means of small sample inference with dichotomous data is Fisher's exact test. However, this method is limited to single-stage designs when there are multiple arms. Therefore, here, we propose a two-stage version of Fisher's exact test, with the potential to stop early to accept or reject null hypotheses, which is applicable to multi-arm studies. In particular, we provide precise formulae describing the requirements for achieving weak or strong control of the familywise error-rate with this design. Following this, we describe how the design parameters may be optimised to confer desirable operating characteristics. For a motivating example based on a phase II clinical trial, we demonstrate that on average our approach is less conservative than corresponding optimal designs based on exact binomial tests.\\


\begin{bottompar}
	\noindent Address correspondence to M. J. Grayling, MRC Biostatistics Unit, Forvie Site, Robinson Way, Cambridge CB2 0SR, UK; Fax: +44-(0)1223-330365; E-mail: mjg211@cam.ac.uk. 
\end{bottompar}

\section{Introduction}
\label{s:intro}

When designing a study, it is not uncommon for the outcome variable of interest to be dichotomous. Unsurprisingly therefore, there is a long history of publications pertaining to the design of studies that compare two binomial proportions. Whilst a normal approximation could be utilised for comparing proportions, in the case of small samples this can lead to substantial inflation of the familywise error-rate (FWER). Therefore, exact statistical methods are required, and one classical approach to this is Fisher's exact test (Fisher, 1935), sometimes also credited to Yates (1934) or Irwin (1935). Unfortunately, prior to the advent of modern computing power this method was prohibitively computationally intensive, and therefore the early literature contains several proposals for approximating the problem of sample size determination (see, for example, Casagrande et al. (1978), Fleiss et al. (1980), and Ury and Fleiss (1980)). Today, computational speed is no longer an issue, and exact methods are readily available, not only for two-arm, but also multi-arm studies (Mehta and Patel, 1983; Mehta and Patel, 1986). Thus, along with alternative exact methods, such as Barnard's test (Barnard, 1945), Fisher's exact test allows for the effectual design and analysis of small fixed-sample studies with binary outcome variables.

Since Wald published his work on the sequential probability ratio test (Wald, 1945), there has been substantial interest in study designs that allow for the interim assessment of hypotheses. Recently, for logistical reasons, the focus of such research has typically been on group, rather than fully, sequential designs. Indeed, much methodology now exists on this (see, for example, Jennison and Turnbull (2000)), including recent extensions to multi-arm studies with normally distributed outcomes (Magirr et al., 2012). This approach has also been demonstrated to be applicable asymptotically to binary outcome variables (Jaki and Magirr, 2013).

Unfortunately, the drawbacks of utilising a normal approximation in small sample studies with binary data persist in this sequential setting. Accordingly, recent proposals have included methods for the design of two-stage multi-arm studiess, allowing early stopping to accept null hypotheses, that employ exact binomial tests (Jung, 2008). Later, this methodology was extended in the case of two-arm designs to allow early stopping to reject null hypotheses (Jung, 2013). However, this approach was demonstrated to be highly conservative for certain values of the shared success probability across the design arms. To attempt to resolve this in a two-arm setting, a recent paper presented a two-stage version of Fisher's exact test (Jung and Sargent, 2014). It was established that this design could on average more exhaustively utilise the designated FWER.

Nonetheless, as stated, the design presented in Jung and Sargent (2014) was limited to two-arm studies only. Moreover, a functional form for the stopping boundaries at the interim analysis was assumed, which may substantially reduce the chance the design will be more efficient than an approach based on exact binomial tests. In this paper, we seek to address these limitations by presenting an optimised two-stage version of Fisher's exact test applicable to multi-arm studies. Extension to a multi-arm domain is of particular importance because of the noted advantages in terms of efficiency of comparing multiple arms against a single control (Parmar et al., 2014). Furthermore, this study design scenario is becoming increasingly common. For example, in clinical research, in many disease settings numerous novel compounds are available for testing in phase II, and the primary outcome variable is typically a binary response indicator. Whilst randomised trials are now being far more typically utilised in this setting (Ivanova et al., 2016).

This paper will consist of the following sections. First we introduce the notation used in the paper in Section 2.1. Following this, in Section 2.2, we summarise the previously proposed design based on exact binomial tests. Our approach is then presented in Section 2.3. Throughout these sections, particular focus is given to the requirements for achieving weak or strong control of the FWER. Then, in Section 2.4 we introduce a phase II clinical trial utilised as a motivating example. Our results are then detailed in Section 3, before we conclude with a discussion of the advantages and disadvantages of the two considered designs in Section 4.

\section{Methods} \label{s:model}

\subsection{Notation, hypotheses and analysis}

Define the sets $\mathbb{N}_a = \{n\in\mathbb{N} : n\le a\}$ and $\mathbb{N}_a^+=\mathbb{N}_a\backslash\{0\}$. Then, we suppose that our study will have a single (control) arm, indexed by $k=0$, which will be compared to $K\in\mathbb{N}^+$ (experimental) arms, indexed by $k\in\mathbb{N}_K^+$, in a pairwise manner within a two-stage design. Denoting the success probability for each arm $k\in\mathbb{N}_K$ by $p_k$, we test the following composite hypotheses

\[ H_{0k} : p_k=p_0, \qquad H_{1k} : p_k=p_0+\delta_k, \qquad k\in\mathbb{N}_K^+, \]

where $\delta_k>0$ for $k\in\mathbb{N}_K^+$. That is, if $Y_{ki}\in\mathbb{N}_1$ is the random variable describing the outcome from experiment $i$ in arm $k$, then $\mathbb{P}(Y_{ki}=1)=p_k$ for $k\in\mathbb{N}_K^+$.

We allow early stopping to both reject and accept the null hypotheses, if so desired, assuming that the rejection of any null hypothesis at the first analysis leads to the termination of the whole study. Note that methodology for a design which terminates the study only when a decision has been made for every null hypothesis could be specified similarly.

We assume that control of the FWER, the probability of one or more incorrect rejections of null hypotheses, is desired to some maximal level $\alpha\in(0,1)$. We discuss criteria for both weak and strong control of the FWER. Moreover, we suppose the experiment must have a familywise power (FWP) of at least $1-\beta\in(0,1)$ when $p_k=p_0+\delta_k$ for $k\in\mathbb{N}_K^+$. That is, we design the experiment to have power at reject at least one of the $H_{0k}$, when $H_{1k}$ is true for $k\in\mathbb{N}_K^+$. Note however that power to reject a particular null hypothesis could be achieved similarly.
 
In testing our hypotheses, we will make repeated reference to the odds ratios $\theta_k$ given by

\[ \theta_k=\frac{p_k(1-p_0)}{p_0(1-p_k)}, \qquad k\in\mathbb{N}_K^+.\]

Furthermore, we will make use of the vectors $\boldsymbol{p}=(p_0,\dots,p_K)^\top$, $\boldsymbol{\theta}=(\theta_1,\dots,\theta_K)^\top$, and $\boldsymbol{\delta}=(0,\delta_1,\dots,\delta_K)^T$. To indicate the $\boldsymbol{\theta}$ implied by a particular $\boldsymbol{p}$ using the relationship given above, we use the notation $\boldsymbol{\theta}(\boldsymbol{p})$.

In stage one, we suppose that $n$ experiments are to be conducted in the control arm, for some $n\in\mathbb{N}^+$. Then, we specify parameters $r_{Cj}\in\mathbb{R}^+\cup\{0\}$ such that the total number of experiments conducted in the control arm by the completion of stage $j\in\mathbb{N}_2$ is $r_{Cj}n\in\mathbb{N}$. Similarly, we designate values $r_{Ej}\in\mathbb{R}^+\cup\{0\}$ such that the total number of experiments conducted in each arm $k\in\mathbb{N}_K^+$ still present in the experiment is $r_{Ej}n\in\mathbb{N}$. Here, we retain the notion of a stage 0 to simplify the expressions that follow. To facilitate the dropping of one or more arms after stage one, we denote by $r_{kj}n\in\mathbb{N}$ the actual number of trials conducted in arm $k\in\mathbb{N}_K$ after stage $j\in\mathbb{N}_2$. Note therefore that $r_{C0}=r_{E0}=0$ and $r_{C1}=1$, the other $r_{Cj}$ and $r_{Ej}$ will be assumed to be pre-specified. In contrast, $n$ will be determined based on the studies' power requirements. One could however choose to search numerically for advantageous values of the $r_{Cj}$ and $r_{Ej}$, according to some designated optimality criteria, if desired.

Next, we denote by $X_{kj}$ the unknown total number of successes in arm $k\in\mathbb{N}_K$ in stage $j\in\mathbb{N}_2^+$, and by $x_{kj}$ its corresponding observed value. Thus $X_{kj}\sim B\{n(r_{kj}-r_{kj-1}),p_k\}$.

Then, in both the Fisher exact and exact binomial test frameworks, at each interim analysis $j\in\mathbb{N}_2^+$, we employ the following test statistics

\begin{equation}\label{test}
T_{kj}=\sum_{m=1}^jx_{km} - \sum_{m=1}^jx_{0m}, \qquad k\in\mathbb{N}_K^+.
\end{equation}

Finally, in what follows it will be convenient to formalise the eventual (unknown) outcome of the study. We achieve this via the pairs $(\boldsymbol{\Psi},\boldsymbol{\Omega})$, where \(\boldsymbol{\Psi} = (\Psi_1,\dots,\Psi_K)^\top\) and \(\boldsymbol{\Omega} = (\Omega_1,\dots,\Omega_K)^\top\) with
\begin{itemize}
	\item \(\Psi_k \in \{0,1\}\), with \(\Psi_{k}=1\) if \(H_{0k}\) is rejected, and \(\Psi_{k}=0\) otherwise,
	\item \(\Omega_{k} \in \{1,2\}\), with \(\Omega_{k}=1\) if following stage one $H_{0k}$ is rejected or accepted, or the whole study is stopped and no decision is made on $H_{0k}$, and \(\Omega_{k}=2\) otherwise.
\end{itemize}

Using this notation, recalling our previous prescription that the rejection of one or more null hypotheses at the interim analysis causes the termination of the entire study, we can define the sample space of the random pairs $(\boldsymbol{\Psi},\boldsymbol{\Omega})$ by
\begin{align*}
\Xi &=\left\{\vphantom{\sum_{k=1}^{K}\mathbb{I}(\omega_k > j)}(\boldsymbol{\psi},\boldsymbol{\omega})\in\{0,1\}^{K}\times\{1,2\}^{K} : \text{if } \sum_{k=1}^{K}\mathbb{I}(\omega_k = 1)\mathbb{I}(\psi_k=1) \ge 1 \text{ then } \sum_{k=1}^{K}\mathbb{I}(\omega_k = 2) = 0 \right\}.
\end{align*}
Here, $\mathbb{I}(A)$ is the indicator function on event $A$. The following sets will now also be useful
\begin{align*}
\Xi_{\text{ind}}(k) &= \left\{(\boldsymbol{\psi},\boldsymbol{\omega})\in\Xi : \mathbb{I}(\psi_k=1) \right\},\\
\Xi_{\text{rej}} &= \left\{(\boldsymbol{\psi},\boldsymbol{\omega})\in\Xi : \sum_{k=1}^K\mathbb{I}(\psi_k=1)>0 \right\},\\
\Xi_{\text{FWER}}(\boldsymbol{p}) &= \left\{(\boldsymbol{\psi},\boldsymbol{\omega})\in\Xi : \sum_{k=1}^K\mathbb{I}(\psi_k=1)\mathbb{I}(p_k=p_0)>0 \right\}.
\end{align*}

These sets represent the subset of outcomes in which a single particular null hypothesis, at least one null hypothesis, or at least one true null hypothesis, is rejected respectively.

Then, it is the ability to evaluate the probability of observing $(\boldsymbol{\Psi},\boldsymbol{\Omega})=(\boldsymbol{\psi},\boldsymbol{\omega})$ on trial completion, for any vector of success probabilities $\boldsymbol{p}$, that is key to determining and optimising the considered designs. Explicitly, referring to this probability as $\mathbb{P}(\boldsymbol{\psi},\boldsymbol{\omega}|\boldsymbol{p})$, we have
\begin{align}
\mathbb{P}(\text{Reject } H_{0k} \mid \boldsymbol{p}) &= \sum_{(\boldsymbol{\psi},\boldsymbol{\omega})\in\Xi_{\text{ind}(k)}}\mathbb{P}(\boldsymbol{\psi},\boldsymbol{\omega}|\boldsymbol{p}),\\
	FWP(\boldsymbol{p}) &= \sum_{(\boldsymbol{\psi},\boldsymbol{\omega})\in\Xi_{\text{rej}}}\mathbb{P}(\boldsymbol{\psi},\boldsymbol{\omega}|\boldsymbol{p}),\\
		FWER(\boldsymbol{p}) &= \sum_{(\boldsymbol{\psi},\boldsymbol{\omega})\in\Xi_{\text{FWER}(\boldsymbol{p})}}\mathbb{P}(\boldsymbol{\psi},\boldsymbol{\omega}|\boldsymbol{p}),\\
			ESS(\boldsymbol{p}) &= \sum_{(\boldsymbol{\psi},\boldsymbol{\omega})\in\Xi}n\left(r_{C \max_k \omega_k} + \sum_{k=1}^K r_{E\omega_k} \right)\mathbb{P}(\boldsymbol{\psi},\boldsymbol{\omega}|\boldsymbol{p}),
\end{align}
where $FWP$, $FWER$ and $ESS$ are functions that evaluate the FWP, FWER and expected sample size (ESS) for a given $\boldsymbol{p}$.

\subsection{Exact binomial testing design}

In this section, we discuss design based upon exact binomial tests. Jung (2008) and Jung (2013) together provide detailed discussions on such designs in the case $K=1$, and some guidance for $K>1$. We expand on these considerations, providing formulae for computing the FWER, FWP, and ESS of any design, with any value for $K$, and early stopping to reject and accept null hypotheses as desired.

In this approach, the goal is to identify suitable values for the parameter $n\in\mathbb{N}^+$ discussed earlier, along with acceptance and rejection boundaries $\boldsymbol{f}=(f_1,f_2)^\top$ and $\boldsymbol{e}=(e_1,e_2)^\top$ respectively. Informally, if $T_{kj}\le f_j$ we accept $H_{0k}$, whilst $T_{kj}\ge e_j$ results in the rejection of $H_{0k}$. The space of possible designs is given by
\begin{align*}
\mathscr{D} &= \left\{ (n,\boldsymbol{f},\boldsymbol{e})\in\mathbb{N}_{n_{\text{max}}}\backslash\mathbb{N}_{n_{\text{min}}}\times\mathbb{Z}^2\times\mathbb{Z}^2 : r_{C2}n,\ r_{E1}n,\ r_{E2}n\in\mathbb{N}^+,\ e_2=f_2+1,\ f_1<e_1-1, \right.\\
& \qquad \qquad \left. f_1\in\{-r_{C1}n,\dots,r_{E1}n-2\}, \ e_1\in\{-r_{C1}n+2,\dots,r_{E1}n\},\right.\\
& \qquad \qquad \qquad \qquad \left. f_2\in\{f_1+1-(r_{C2}-r_{C1})n,\dots,e_1-1+(r_{E2}-r_{E1})n\} \right\}.
\end{align*}
Here, $(n_{\text{min}},n_{\text{max}})\in\{(p,q)\in\mathbb{N}^+\times\mathbb{N}^+ : p \le q\}$, place logical limits on the allowed value of $n$. These could be chosen, for example, based on the sample size required by a corresponding single-stage design. Moreover, the restrictions ensure for any $n\in\mathbb{N}^+$ that it is possible to stop to accept or reject null hypotheses at both analyses, and ensure the study terminates after at most two stages. Note that if it desired to prevent the possibility to reject or accept null hypotheses at the end of stage one, further restrictions can simply be placed on the set $\mathscr{D}$, with design optimisation then proceeding as below.

Having specified $n$, $\boldsymbol{f}$ and $\boldsymbol{e}$, the studies' formal conduct can be defined, along with the formulae for $\mathbb{P}(\boldsymbol{\psi},\boldsymbol{\omega}|\boldsymbol{p})$. We provide both in the Appendix. Here, we proceed directly to discussing how values for $n$, $\boldsymbol{a}$ and $\boldsymbol{r}$ can be chosen. Specifically, in this instance, it was proposed by Jung (2008) that an optimal design be determined by exhaustively searching over the set $\mathscr{D}$. Since the evaluation of each design is independent, parallel execution can be used to enhance the speed of this search. Extending Jung (2013), we search for the solution to
\begin{align*}
\argmin_{(n,\boldsymbol{a},\boldsymbol{r})\in\mathscr{D}} \ \ &w_1ESS(\boldsymbol{p}_{\text{ESS}}) + w_2ESS(\boldsymbol{p}_{\text{ESS}}+\boldsymbol{\delta}) + w_3n(r_{C2}+Kr_{E2}),\\
		\text{subject to } \qquad \alpha & \ge \max_{\boldsymbol{p}\in \text{P}_{\text{FWER}}}FWER(\boldsymbol{p}),\\
		1 - \beta & \le \min_{p\in [0,1-\max_k \delta_k]}FWP\{(p,\dots,p)^\top+\boldsymbol{\delta}\},
\end{align*}
where $w_1,w_2,w_3\in\mathbb{R}^+\cup\{0\}$ are weights that indicate which of the three factors we desire to minimise most. Heeding the advice of Mander et al. (2012), we always ensure that $w_1+w_2>0$ since many designs will likely share the same minimal maximal sample size. Later, we will make use of the notation $\boldsymbol{w}=(w_1,w_2,w_3)$. Additionally, $\boldsymbol{p}_{\text{ESS}}$ is a specified vector of success probabilities to utilise in the optimisation procedure. Typically, this will be those expected under a global null hypotheses $p_k=p_0$ for $k\in\mathbb{N}_K^+$. The two given constraints here are our requirements on the studies' FWER and FWP. The exact form of the set $\text{P}_{\text{FWER}}$ is dependent on whether weak or strong control of the FWER is desired. In the case where we designate that weak control (i.e., control when all null hypotheses are true) must be achieved, we set $\text{P}_{\text{FWER}} = \{\boldsymbol{p}\in[0,1]^{K+1} : \boldsymbol{p}=(p,\dots,p)^\top,\ p\in[0,1] \}$. Alternatively, for strong control we must take $\text{P}_{\text{FWER}} = [0,1]^{K+1}$.

Note that this contradicts the advice given in Jung (2008), which stated that strong control could be achieved for the case of two-arm studies by controlling the FWER for $\boldsymbol{p}=(0.5,0.5)^\top$. Although the true maximal FWER appears in general to be close to that attained when $\boldsymbol{p}=(0.5,0.5)^\top$, a simple counter example to it being universally true can be constructed by considering a design with $a_1,a_2 \le 0$. With such a design, $\boldsymbol{p}=(0,0)^\top$ attains a larger FWER than $\boldsymbol{p}=(0.5,0.5)^\top$. Consequently, a search over one of the specified $\text{P}_{\text{FWER}}$ must be included when determining a design. In practice, the multi-dimensional search required for strong control must be broached by utilising the criteria for weak control during the design determination stage. Then after choosing an optimal design, a retrospective search over $\boldsymbol{p}\in[0,1]^{K+1}$ should be performed to argue that strong control has been achieved. Intuitively, it is logical that the maxima will occur when $\boldsymbol{p}\in\{\boldsymbol{q}\in[0,1]^{K+1} : \boldsymbol{q}=(q,\dots,q)^\top,\ q\in[0,1] \}$. This type of problem is common to experiments with binary outcome variables (see, for example, Kunzmann and Kieser (2016)).

With this, the methodology required for determining an optimal design based on exact binomial tests has been specified. We will later compare this approach to our two-stage Fisher exact test.

\subsection{Fisher exact test design}

As was discussed, the exact binomial test method summarised above was confirmed to be highly conservative for many combinations of success probabilities, and to address this in a two-arm setting Jung and Sargent (2014) proposed a two-stage version of Fisher's exact test. In this section, we detail our extension to their proposal; allowing for multiple arms and the optimisation of the stopping boundaries.

Our test here is based on the conditional distribution of the $T_{kl}$, specified earlier in Equation~(\ref{test}), given the observed total number of successes in each completed stage $j\in\mathbb{N}_2^+$, $z_j=x_{0j}+\dots+x_{Kj}$ (with the unknown total number of stage-wise successes denoted by $Z_j$). To achieve this, we let $\boldsymbol{\rho}_j=(\rho_{1j},\dots,\rho_{Kj})^\top$, with $\rho_{kj}=1$ if arm $k$ is present in the study in stage $j$, and  $\rho_{kj}=0$ otherwise. Therefore, note that $\boldsymbol{\rho}_1=\boldsymbol{1}$. Then, extending Jung and Sargent (2014), the probability mass function of $\boldsymbol{x}_j=(x_{0j},\dots,x_{Kj})^\top$ conditional on $z_j$, $\boldsymbol{\rho}_j$ and $\boldsymbol{\theta}$ is
\[ f(\boldsymbol{x}_j\mid z_j,\boldsymbol{\rho}_j,\boldsymbol{\theta}) = \frac{\mathbb{I}\left(\sum_{k=0}^Kx_{kj} = z_j\right)h(\boldsymbol{x}_j)}{\sum_{a_{0j}=0}^{(r_{Cj} - r_{Cj-1})n}\sum_{a_{1j}=0}^{(r_{Ej} - r_{Ej-1})n}\dots\sum_{a_{Kj}=0}^{(r_{Ej} - r_{Ej-1})n}\mathbb{I}\left(\sum_{k=0}^K a_{kj} = z_j\right)h(\boldsymbol{a}_j)} ,\]
where
\[ h(\boldsymbol{x}_j) = \begin{pmatrix} (r_{Cj} - r_{Cj-1})n \\ x_{0l} \end{pmatrix} \prod_{k=1}^K \begin{pmatrix} \rho_{kj}(r_{Ej} - r_{Ej-1})n \\ x_{kj} \end{pmatrix}\theta_k^{x_{kj}},\]
and we set $^0C_x = \mathbb{I}\{x=0\}$. Note that this immediately implies $f(\boldsymbol{x}_j\mid z_j,\boldsymbol{\rho}_j,\boldsymbol{\theta})=0$ if
\[ \sum_{k=1}^K \mathbb{I}(\rho_{kj} = 0)\mathbb{I}(x_{kj} > 0)>0. \]
Now, the studies' conduct depends upon having chosen stopping boundaries for all possible total number of successes that could be observed in stage one and stage two, for every possible number of experimental arms that could be present in stage two. Formally, the following are required 
\begin{itemize}
	\item $e_{1z_1}$, $f_1$ (with $e_{1z_1}>f_1+1$) for $z_1=0,\dots,(r_{C1}+Kr_{E1})n$ and;
	\item $e_{2kz_2z_1}=f_{2kz_2z_1}+1$ for $k\in\mathbb{N}_K^+$, $z_1=0,\dots,(r_{C1}+Kr_{E1})n$, $z_2=0,\dots,\{(r_{C2}-r_{C1})+k(r_{E2}-r_{E1})\}n$.
\end{itemize}
Thus, whilst the rejection boundary in stage one depends on the number of observed successes, we have chosen to simplify matters by making $f_1$ independent of $z_1$, as in Jung and Sargent (2014).

The studies' formal conduct is then as follows
\begin{enumerate}
	\item Set $\boldsymbol{\psi}=\boldsymbol{\omega}=\boldsymbol{0}$.
	\item Conduct stage one of the study, allocating $r_{C1}n$ patients to the control arm, and $r_{E1}n$ patients to each arm $k\in\mathbb{N}_K^+$. Following data accrual, compute $z_1$ and the $T_{k1}$.
	\item For each $k\in\mathbb{N}_K^+$
	\begin{itemize}
		\item If $T_{k1}\ge e_{1z_1}$ reject $H_{0k}$, setting $\psi_k=1$ and $\omega_k=1$.
		\item If $T_{k1}\le f_1$ accept $H_{0k}$, setting $\omega_k=1$.
	\end{itemize}
	\item If $\sum_{k=1}^K\mathbb{I}(\psi_k=1)=0$ and $\sum_{k=1}^K\mathbb{I}(\omega_k=0)>0$, continue to 5. Otherwise stop the study, and for each $k\in\mathbb{N}_K^+$ with $\omega_k=0$, set $\omega_k=1$.
	\item Set $\boldsymbol{\rho}_2=\{\mathbb{I}(\omega_1=0),\dots,\mathbb{I}(\omega_K=0)\}^\top$.
	\item Conduct stage two of the study, allocating $(r_{C2}-r_{C1})n$ patients to the control arm, and $(r_{E2}-r_{E1})n$ patients to each arm $k\in\mathbb{N}_K^+$ with $\rho_{k2}=1$. Following data accrual, compute $z_2$ and the $T_{k2}$.
	\item For each $k\in\mathbb{N}_K^+$ with $\rho_{k2}=1$
	\begin{itemize}
		\item If $T_{k2}\ge e_{2\boldsymbol{\rho}_2\cdot\boldsymbol{\rho}_2z_2z_1}$ reject $H_{0k}$, setting $\psi_k=1$ and $\omega_k=2$.
		\item If $T_{k2}\le f_{2\boldsymbol{\rho}_2\cdot\boldsymbol{\rho}_2z_2z_1}$ accept $H_{0k}$, setting $\omega_k=2$.
	\end{itemize}
\end{enumerate}

The above specifies the conduct of our study given values for $n$, the allocation ratios $r_{Cl}$ and $r_{El}$, and the required stopping boundaries. At the design stage of a study though, we require the ability to choose suitable values for $n$ and the stopping boundaries. The large number of required stopping boundaries precludes the possibility of optimising every chosen value, as in the method of the previous section. The aim of the Fisher exact approach though is not to optimise every boundary, but to instead ensure conditional control of the FWER for all possible values of $z_1$, $z_2$ and $\boldsymbol{\rho}_2$, such that marginal control is then certain. 

However, we can identify designs with more desirable operating characteristics. In Jung (2013) and Jung and Sargent (2014), the stopping boundaries for stage one were pre-specified, such that only those for the second stage needed to be chosen. Designing a study in this manner reduces the computational complexity of the ensuant optimisation problem. However, it reduces the chance that the operating characteristics of the resultant design will compare favourably with a design determined using the exact binomial test approach. Consequently, we here propose a more flexible design framework, reliant on the specification of two parameters, $\alpha_1\in(0,\alpha)$ and $\beta_1\in(0,\beta)$, that can then be optimised. We first describe how the stopping boundaries are chosen for any $n$, given $\alpha_1$ and $\beta_1$. Following this, we detail how $n$ can be chosen such that the study attains the desired FWP.

First, for any $n$, define the function $\alpha_{I1}(z_1\mid r_{1z_1},\boldsymbol{\theta})$ as follows. Precisely, this describes the probability of committing a familywise error at the end of stage one if $z_1$ success are observed, a rejection boundary of $e_{1z_1}$ is utilised, and the true vector of odds ratios is $\boldsymbol{\theta}$. We have

\[ \alpha_{I1}(z_1\mid r_{1z_1},\boldsymbol{\theta}) = \sum_{x_{01}=0}^{r_{C1}n}\sum_{x_{11}=0}^{r_{E1}n}\dots\sum_{x_{K1}=0}^{r_{E1}n}\mathbb{I}\left\{ \sum_{k=1}^K\mathbb{I}(T_{k1}\ge e_{1z_1})\mathbb{I}(\theta_k=1)>0 \right\}f(\boldsymbol{x}_1\mid z_1,\boldsymbol{1},\boldsymbol{\theta}). \]
Motivated by the error spending approach to group sequential trial design, when desiring weak control of the FWER, we select $e_{1z_1}$ for $z_1\in\{0,\dots,(r_{C1}+Kr_{E1})n\}$ as the minimal integer for which $\alpha_{I1}(z_1\mid e_{1z_1},\boldsymbol{1}) \le \alpha_1$. Alternatively, for strong control $e_{1z_1}$ is instead the smallest integer such that
\[ \max_{\boldsymbol{\theta}\in(0,\infty)^{K}}\alpha_{I1}(z_1\mid e_{1z_1},\boldsymbol{\theta}) \le \alpha_1. \]
That is, we either weakly or strongly control the possibility of committing a familywise error at the first analysis to $\alpha_1$.

Next, define
\[ \beta_{II1}(z_1|f_1,\boldsymbol{\theta}) = \sum_{x_{01}=0}^{r_{C1}n}\sum_{x_{11}=0}^{r_{E1}n}\dots\sum_{x_{K1}=0}^{r_{E1}n}\mathbb{I}(T_{11}\le f_{1})f(\boldsymbol{x}_1|z_1,\boldsymbol{1},\boldsymbol{\theta}).\]
We then choose $f_1$ to be the largest integer such that the marginal type-II error rate for $H_{01}$ at the first stage, when $\boldsymbol{p}=(p,\dots,p)^\top+\boldsymbol{\delta}$, is at most $\beta_1$. That is, $f_1$ is the largest integer such that
\[ \max_{p\in[0,1-\max_k\delta_k]}\sum_{z_1=0}^{(r_{C1}+Kr_{E1})n} \beta_{II1}[z_1|f_1,\boldsymbol{\theta}\{(p,\dots,p)^\top+\boldsymbol{\delta}\}]g\{z_1\mid(p,\dots,p)^\top+\boldsymbol{\delta},\boldsymbol{1}\} \le \beta_1. \]
Here $g(z_j |\boldsymbol{\rho}_j,\boldsymbol{p})$ is the probability mass function of $Z_j$ given $\boldsymbol{\rho}_j$ and $\boldsymbol{p}$
\begin{align*}
g(z_j  | \boldsymbol{p},\boldsymbol{\rho}_j) &= \sum_{x_{0j}=0}^{(r_{Cj}-r_{Cj-1})n}\sum_{x_{1j}=0}^{(r_{Ej}-r_{Ej-1})n}\dots\sum_{x_{Kj}=0}^{(r_{Ej}-r_{Ej-1})n}\mathbb{I}\left\{ \sum_{k=0}^Kx_{kj}=z_j \right\}\\
& \qquad \qquad \qquad b\{x_{0l}|(r_{Cj}-r_{Cj-1})n,p_0\}\prod_{k=1}^Kb\{x_{kl}|\rho_{kj}(r_{Ej}-r_{Ej-1})n,p_k\},
\end{align*}
with $b(x|n,p)= \phantom{.}^nC_xp^x(1-p)^{n-x}$.

Now, define the function $\alpha_{I2}(k,z_2,z_1|e_{2kz_1z_1},f_1,e_{1z_1},\boldsymbol{\theta})$. This evaluates the probability of committing a familywise error at the end of the second stage, if $z_1$ and $z_2$ successes are observed in stages one and two respectively, with $k$ experimental arms present in the second stage, conditional on the use of the stopping boundaries $e_{2kz_1z_1}$, $f_1$, and $e_{1z_1}$, and a nominated value of $\boldsymbol{\theta}$. Precisely
\begin{align*}
\alpha_{I2}(k,z_2,z_1|e_{2kz_1z_1},f_1,e_{1z_1},\boldsymbol{\theta}) &= \sum_{x_{01}=0}^{r_{C1}n}\sum_{x_{11}=0}^{r_{E1}n}\dots\sum_{x_{K1}=0}^{r_{E1}n}\sum_{x_{02}=0}^{(r_{C2}-r_{C1})n}\sum_{x_{12}=0}^{(r_{E2}-r_{E1})n}\dots\\
& \qquad \sum_{x_{K2}=0}^{(r_{E2}-r_{E1})n} \left( \mathbb{I}\left[ \sum_{k_1=1}^K\mathbb{I}\{T_{k_11}\in(f_1,e_{1z_1})\} = k \right] \right)\left\{ \prod_{k_2=1}^K\mathbb{I}(T_{k_21}<e_{1z_1}) \right\}\\
& \qquad \qquad \left( \mathbb{I}\left[ \sum_{k_3=1}^K\mathbb{I}\{T_{k_31}\in(f_1,e_{1z_1})\}\mathbb{I}\{T_{k_32}\ge e_{2kz_2z_1}\}\mathbb{I}\{\theta_k=1\}>0 \right] \right)\\
& \qquad \qquad \qquad f(\boldsymbol{x}_1|z_1,\boldsymbol{1},\boldsymbol{\theta})f(\boldsymbol{x}_2|z_2,\boldsymbol{\rho}_2,\boldsymbol{\theta}),
\end{align*}
where $\boldsymbol{\rho}_2$ can be computed from the $x_{k1}$ and the stopping rules.

Then, to ensure weak control of the FWER we choose $e_{2kz_2z_1}$ for $k=1,\dots,K$, $z_1=0,\dots,(r_{C1}+Kr_{E1})n$, $z_2=0,\dots,[(r_{C2}-r_{C1})+k(r_{E2}-r_{E1})]n$ to be the smallest integer such that
\[ \alpha_{I2}(k,z_2,z_1|e_{2kz_1z_1},f_1,e_{1z_1},\boldsymbol{1}) \le \frac{\alpha-\alpha_{I1}(z_1\mid e_{1z_1},\boldsymbol{1})}{K}. \]
Alternatively, for strong control we must ensure that
\[ \max_{\boldsymbol{\theta}\in(0,\infty)^{K}} \alpha_{I2}(k,z_2,z_1|e_{2kz_1z_1},f_1,e_{1z_1},\boldsymbol{\theta}) \le \frac{\alpha-\max_{\boldsymbol{\theta}\in(0,\infty)^{K}}\alpha_{I1}(z_1\mid e_{1z_1},\boldsymbol{\theta})}{K}. \]

Here, division by $K$ in the right-hand side of the above formulae is to allocate the unspent familywise error equally across the $K$ scenarios defined by the number of arms remaining in the experiment in stage two.

As in the previous section, for strong control, in practice it is necessary to assume the maximal FWER occurs on the boundary, and then search for the maximal marginal FWER after design determination. Note however that weak control is always assured by this design, unlike for the exact binomial test approach.

We have now fully specified a method for boundary determination given $n$. With the boundaries specified, it is then possible to define the formulae for $\mathbb{P}(\boldsymbol{\psi},\boldsymbol{\omega}|\boldsymbol{p})$ in this design. We provide this in the Appendix. Then, with this formulae ascertained, for any $\alpha_1$ and $\beta_1$, we can ensure our studies' FWP requirement is met by searching for the minimal $n\in\mathbb{N}^+$ such that
\[ 1 - \beta \le \min_{p\in [0,1-\max_k \delta_k]}FWP\{(p,\dots,p)^\top+\boldsymbol{\delta}\}. \]

As alluded to previously, we are then free to optimise $\alpha_1$ and $\beta_1$. Following the previous section, we minimise
\begin{align*}
\argmin_{(\alpha_1,\beta_1)\in(0,\alpha)\times(0,\beta)} \ \ &w_1ESS(\boldsymbol{p}_{\text{ESS}}) + w_2ESS(\boldsymbol{p}_{\text{ESS}}+\boldsymbol{\delta}) + w_3n(r_{C2}+Kr_{E2}).
\end{align*}

Here, we do not need to list constraints on the studies' operating characteristics as they assured as part of design determination for any $\alpha_1$ and $\beta_1$.

Now, since $\alpha_1$ and $\beta_1$ are continuous variables, this minima could in theory be ascertained by utilising a conventional greedy optimisation routine. However, there is no guarantee that this would converge to the global optimum in a parameter space that may contain several local optima, given the discrete nature of the stopping boundaries and required value for $n$. One possible solution therefore is to employ a global continuous optimisation algorithm such as simulated annealing, which has previously been used with success in clinical trial design (see, for example, Wason and Jaki (2012) and Chan and Lee (2013)). Such a search is typically a highly time consuming process however. A compromise can be achieved by instead proposing a set of candidate values for $\alpha_1$ and $\beta_1$, $A_1$ and $B_1$ respectively. The efficiency of the designs in $A_1\times B_1$ can then be evaluated in parallel using a grid-search and what is essentially, depending on $|A_1|$ and $|B_1|$, a near globally optimal design obtained. This method is also advantageous in that the solution for different combinations of $w_1$, $w_2$ and $w_3$ can be attained simultaneously. In practice, $A_1$ can be specified by choosing a range of equally spaced values with $\min(A_1)\approx0$ and $\max(A_1)\approx\alpha$, and similarly for $B_1$. This will be our approach in what follows.

\subsection{Example}

In this coming section, we determine several example designs using our two-stage Fisher exact procedure, and compare their performance to those based on the exact binomial method. We reconsider an example motivated by CALGB 50502; a randomised phase II clinical trial that compared the anti-tumour activity of two regimens in patients with relapsed/refractory classical Hodgkin’s lymphoma. With this, we set $\alpha=0.15$, $\beta=0.2$, $\boldsymbol{\delta}=(0,0.15,...,0.15)^\top$, and $\boldsymbol{p}_{ESS}=(0.7,\dots,0.7)^\top$ (Jung, 2008). For simplicity, we limit our focus to designs with $r_{C2}=2r_{C1}=r_{E2}=2r_{E1}=2$.

In determining the designs based on the exact binomial method, we limit the maximal value of $n$ in our search to $0.75n_{\text{fixed}}$, where $n_{\text{fixed}}$ is the group size required by a corresponding single stage design. For the designs utilising the Fisher's exact method, we determine the near optimal design by taking $A_1=\{0.01,0.02,\dots,0.14\}$, $B_1=\{0.01,0.02,\dots,0.19\}$.

Code to replicate our findings is available upon request.

\section{Results}\label{results}

\subsection{Comparison of optimised and non-optimised two-stage Fisher exact designs}

First, we consider the case with $K=1$. Jung and Sargent (2014) designated $f_1=-1$ and $e_{z_11}=\lceil n\delta_1\rceil+1$ for $z_1=0,\dots,2(K + 1)n$ in this instance. Here, we compare the efficiency of their design to several determined using our optimisation procedure. Table 1 presents the ESS for $\boldsymbol{p}=\boldsymbol{p}_{ESS}=(0.7,0.7)^\top$ and $\boldsymbol{p}=\boldsymbol{p}_{ESS}+\boldsymbol{\delta}=(0.7,0.85)^\top$, along with the maximal possible sample size, of several designs. Note that in this instance, the criteria for weak and strong control coincide. Thus strong control is guaranteed by the Fisher exact approach.

It can be seen that the optimised designs allow the ESS under $\boldsymbol{p}_{ESS}$ and $\boldsymbol{p}_{ESS}+\boldsymbol{\delta}$ to be reduced by up to 13.1\% ($\boldsymbol{w}=(1,0,0)$) and 17.5\% ($\boldsymbol{w}=(0,1,0)$) respectively, with only small increases to the maximal possible sample size. Moreover, one of the optimised designs reduces the maximal sample size by 8.3\% ($\boldsymbol{w}=(10^{-5},0,1)$), though this comes at a cost to both $ESS(\boldsymbol{p}_{ESS})$ and $ESS(\boldsymbol{p}_{ESS}+\boldsymbol{\delta})$. A compromise can be achieved between reducing the ESSs and the maximal sample size by taking $\boldsymbol{w}=(1,1,1)$. In fact, the design with $\boldsymbol{w}=(1,1,1)$ performs better than Jung and Sargent's design in terms of $ESS(\boldsymbol{p}_{ESS})$, $ESS(\boldsymbol{p}_{ESS}+\boldsymbol{\delta})$, and $\max N$.

\def\arraystretch{1.2}
\begin{table}[htb]
	\begin{center}
		\caption{A comparison of the design from Jung and Sargent (2014) and several optimised designs for different values of $w_1$, $w_2$ and $w_3$ is shown. Here, $K=1$, $\boldsymbol{p}_{ESS}=(0.7,0.7)^\top$, and $\boldsymbol{\delta}=(0,0.15)^\top$.}
		\begin{tabular}{llrrrrrr}
			\hline
			Design & $\boldsymbol{w}$ & $\alpha_1$ & $\beta_1$ & $n$ & $ESS(\boldsymbol{p}_{ESS})$ & $ESS(\boldsymbol{p}_{ESS}+\boldsymbol{\delta})$ & $\max N$\\
			\hline
			Jung and Sargent & N/A & N/A & N/A & 48 & 145.49 & 146.89 & 192\\
			\hline
			\multirow{7}{*}{Optimised} & $(1,0,0)$ & 0.11 & 0.16 & 52 & 126.49 & 125.42 & 208\\
			\cline{2-8}
			&$(0,1,0)$ & \multirow{2}{*}{0.08} & \multirow{2}{*}{0.17} & \multirow{2}{*}{51} & \multirow{2}{*}{126.54} & \multirow{2}{*}{124.93} & \multirow{2}{*}{204} \\
			&$(1,1,0)$ &  &  &  &  & &\\
			\cline{2-8}
			&$(10^{-5},0,1)$ & 0.01 & 0.01 & 44 & 162.54 & 164.60 & 176\\
			\cline{2-8}
			&$(1,1,1)$ & \multirow{3}{*}{0.04} & \multirow{3}{*}{0.10} & \multirow{3}{*}{46} & \multirow{3}{*}{131.54} & \multirow{3}{*}{131.27} & \multirow{3}{*}{184} \\
			&$(1,0,1)$ &  &  &  &  & &\\
			&$(0,1,1)$ &  &  &  &  & &\\
			\hline
		\end{tabular}
	\end{center}
\end{table}

\subsection{Comparison of two-stage Fisher exact and exact binomial test designs}
	
Next, we compare the performance of our proposed two-stage Fisher exact design to that based on the exact binomial test in the case $K=2$. For each approach, we enforce their respective criteria for weak control of the FWER during design determination.

Searching over $A_1\times B_1$ for the Fisher exact approach, in this case the optimal design for each of $\boldsymbol{w}\in\{(1,0,0),(0,1,0),(10^{-5},0,1),(1,1,0),(1,0,1),(0,1,1),(1,1,1)\}$ is given by the case $(\alpha_1,\beta_1)=(0.07,0.17)$. A summary of this design's operating characteristics, along with that for the corresponding optimal designs using the exact binomial test method, are given in Table 2.
	
\def\arraystretch{1.2}
\begin{table}[htb]
	\begin{center}
		\caption{A summary of several optimal designs for the Fisher and exact binomial test approaches. Here, $K=2$, $\boldsymbol{p}_{ESS}=(0.7,0.7,0.7)^\top$, and $\boldsymbol{\delta}=(0,0.15,0.15)^\top$.}
		\begin{tabular}{lrrrrrrrr}
			\hline
			Design & $\boldsymbol{w}$ & $n$& $f_1$ & $e_1$ & $f_2$ & $ESS(\boldsymbol{p}_{ESS})$ & $ESS(\boldsymbol{p}_{ESS}+\boldsymbol{\delta})$&$\max N$\\
			\hline
			Fisher exact test & All below & 38 & N/A & N/A & N/A & 154.2 & 151.7 & 228 \\
			\hline
			\multirow{7}{*}{Exact binomial test} & $(1,0,0)$ & \multirow{4}{*}{37} & \multirow{4}{*}{2} & \multirow{4}{*}{11} & \multirow{4}{*}{7} & \multirow{4}{*}{144.2} & \multirow{4}{*}{190.3} & \multirow{4}{*}{222}\\
			& $(10^{-5},0,1)$ &  & & &  &  & & \\
			& $(1,0,1)$ &  &  & &  &  & & \\
			& $(1,1,1)$ &  &  & &  &  & & \\
			\cline{2-9}
			& $(0,1,0)$ & 47 & 4 & 8 & 9 & 158.0 & 170.5 & 282 \\
			\cline{2-9}
			& $(1,1,0)$ & 44 & 3 & 8 & 9 & 156.3 & 171.0 & 264 \\
			\cline{2-9}
			& $(0,1,1)$ & 38 & 1 & 9 & 8 & 156.9 & 181.4 & 228 \\
			\hline
		\end{tabular}
	\end{center}
\end{table}

We can see that the Fisher exact design attains the smallest value of $ESS(\boldsymbol{p}_{ESS}+\boldsymbol{\delta})$, being at least 11\% smaller than that for the exact binomial test designs. However, an exact binomial test design does exist with a smaller value for $ESS(\boldsymbol{p}_{ESS})$. The maximal possible required sample size of the Fisher exact design is comparable to that of several of the exact binomial test designs. 

The operating characteristics of these designs are further elucidated in Figure 1, which depicts the FWER and ESS when $\boldsymbol{p}=(p,\dots,p)^\top$ for $p\in[0,1]$, as well as the FWP and ESS when $\boldsymbol{p}=(p,\dots,p)^\top+\boldsymbol{\delta}$ for $p\in[0,0.85]$.

For more extreme values of $p$, the Fisher exact design has a FWER closer to the nominal level. There is however a region around $p=0.5$ in which the exact binomial test designs more exhaustively utilise the allowed FWER. This mirrors the results previously presented for the case with $K=1$.

It is clear that the Fisher exact design almost universally has a larger FWP than the exact binomial test designs. The exact binomial test designs display a characteristic in which the FWP can drop as $p$ tends towards 0.85. This is a consequence of the fact that the probability of observing a large number of successes in arm 0 grows more quickly in this region than that for arms 1 and 2.

With the exception of the optimal exact binomial test design for $\boldsymbol{w}=(1,0,0)$, the designs require similar ESSs when $\boldsymbol{p}=(p,\dots,p)^\top$. When $\boldsymbol{p}=(p,\dots,p)^\top+\boldsymbol{\delta}$ however, the ESS for the Fisher exact design is far smaller than that for all of the exact binomial test designs. The fundamental shape of the ESS curves for the Fisher exact design and the exact binomial test designs in this instance are also different. This reflects the ability of the Fisher exact test design to alter the rejection boundary when a large number of successes are observed.

\begin{figure}
\centering
\subfloat[]{%
	\includegraphics[width=0.5\textwidth]{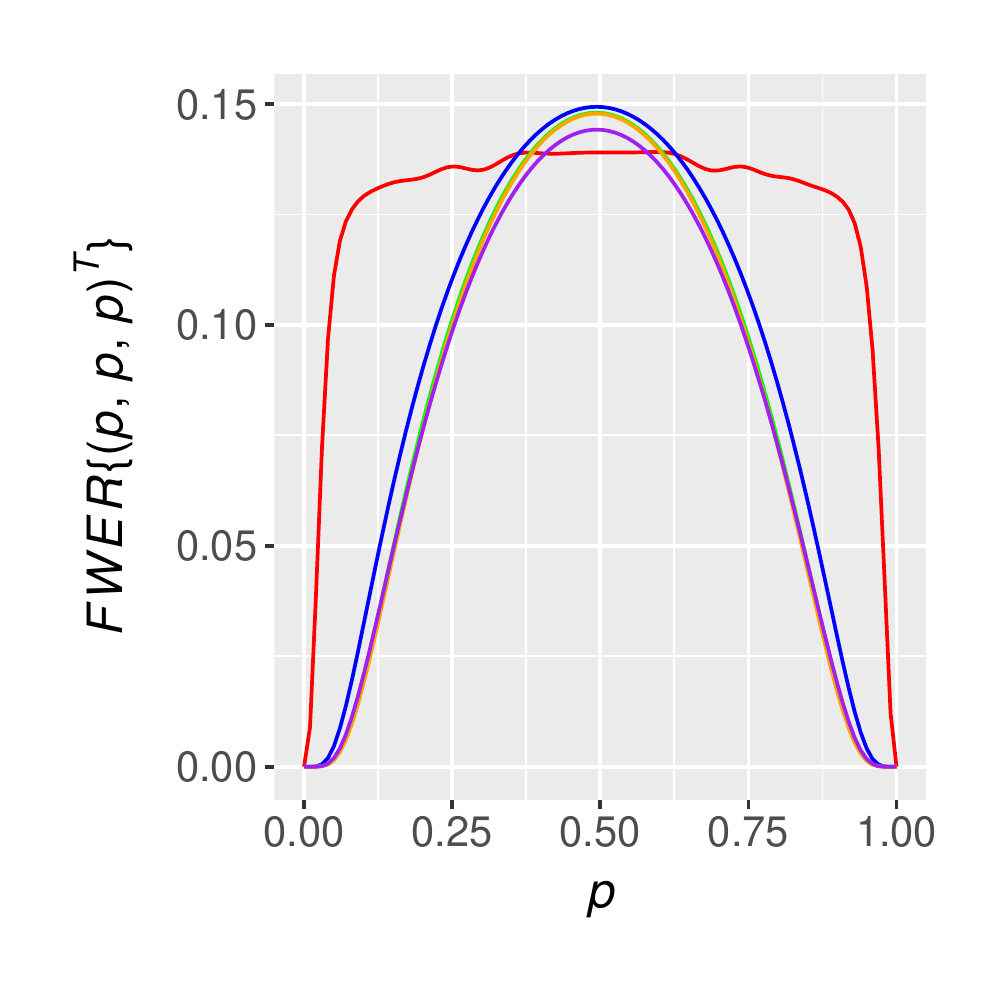}
}
\subfloat[]{%
	\includegraphics[width=0.5\textwidth]{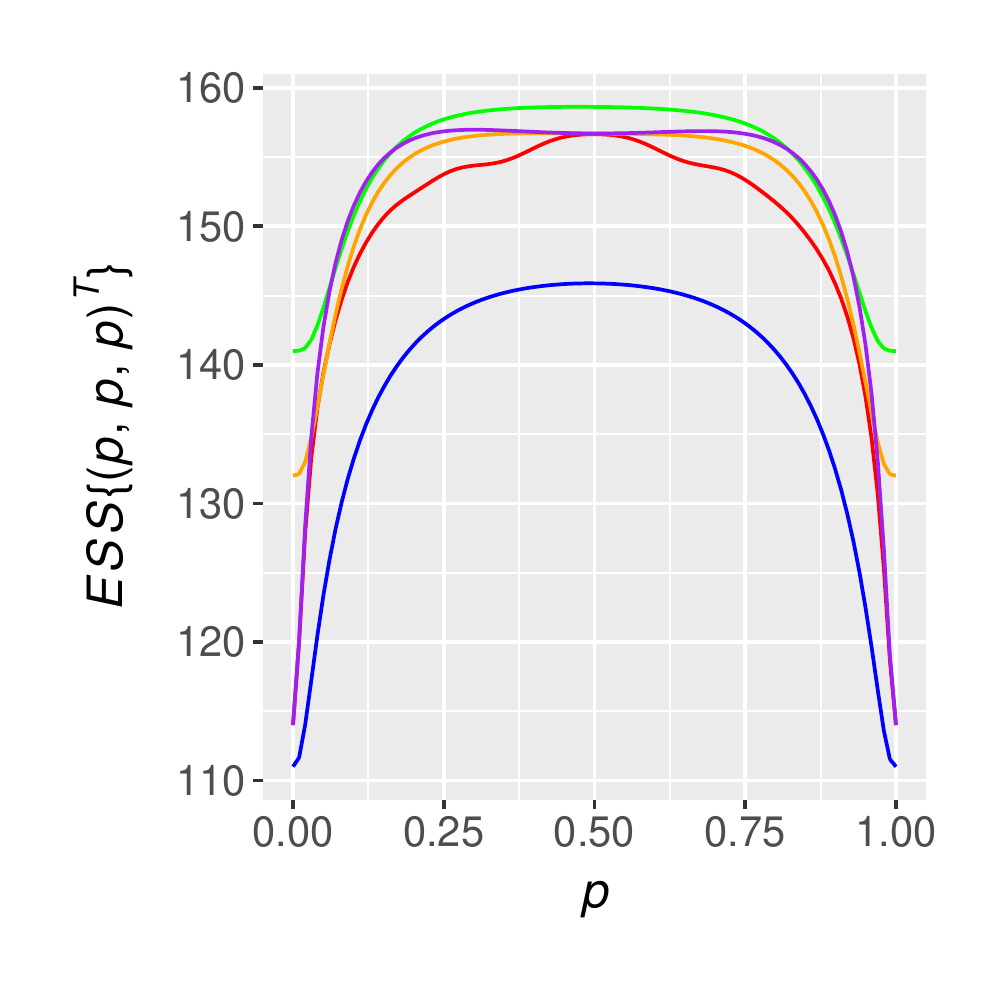}
}

\subfloat[]{%
	\includegraphics[width=0.5\textwidth]{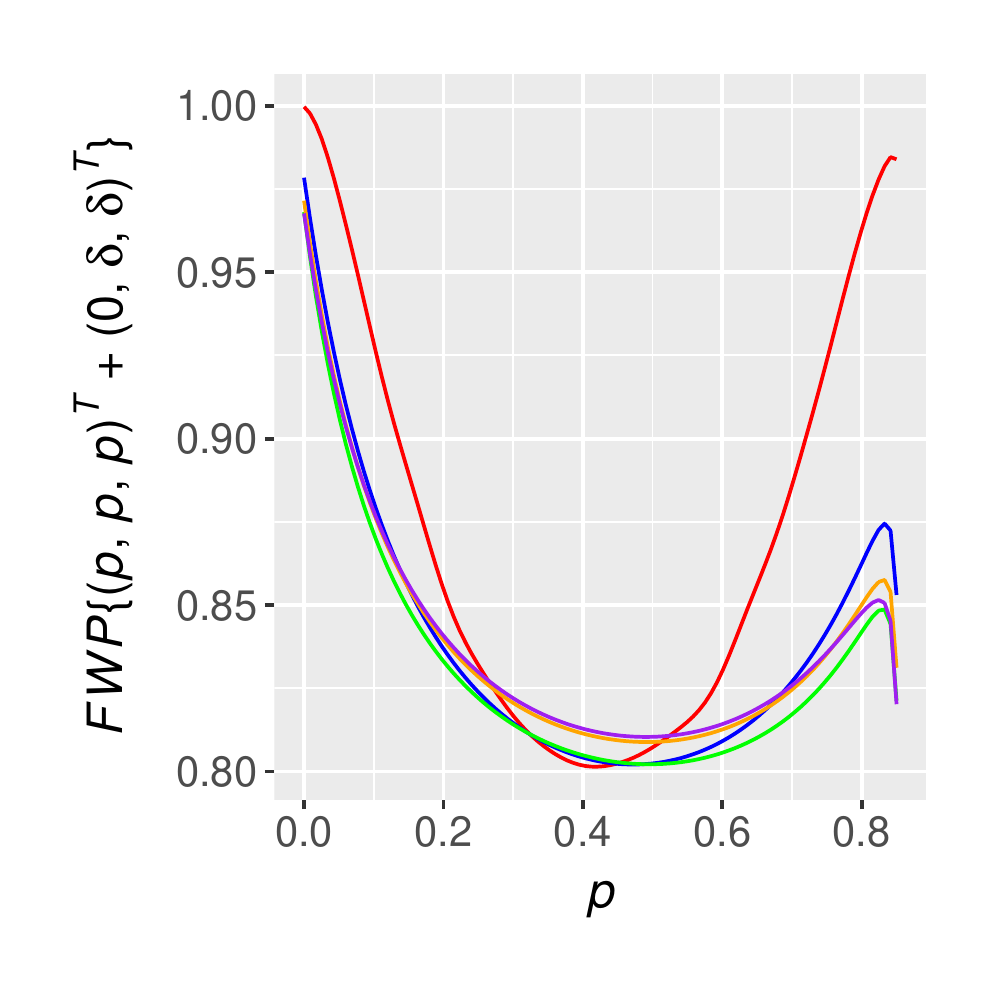}
}
\subfloat[]{%
	\includegraphics[width=0.5\textwidth]{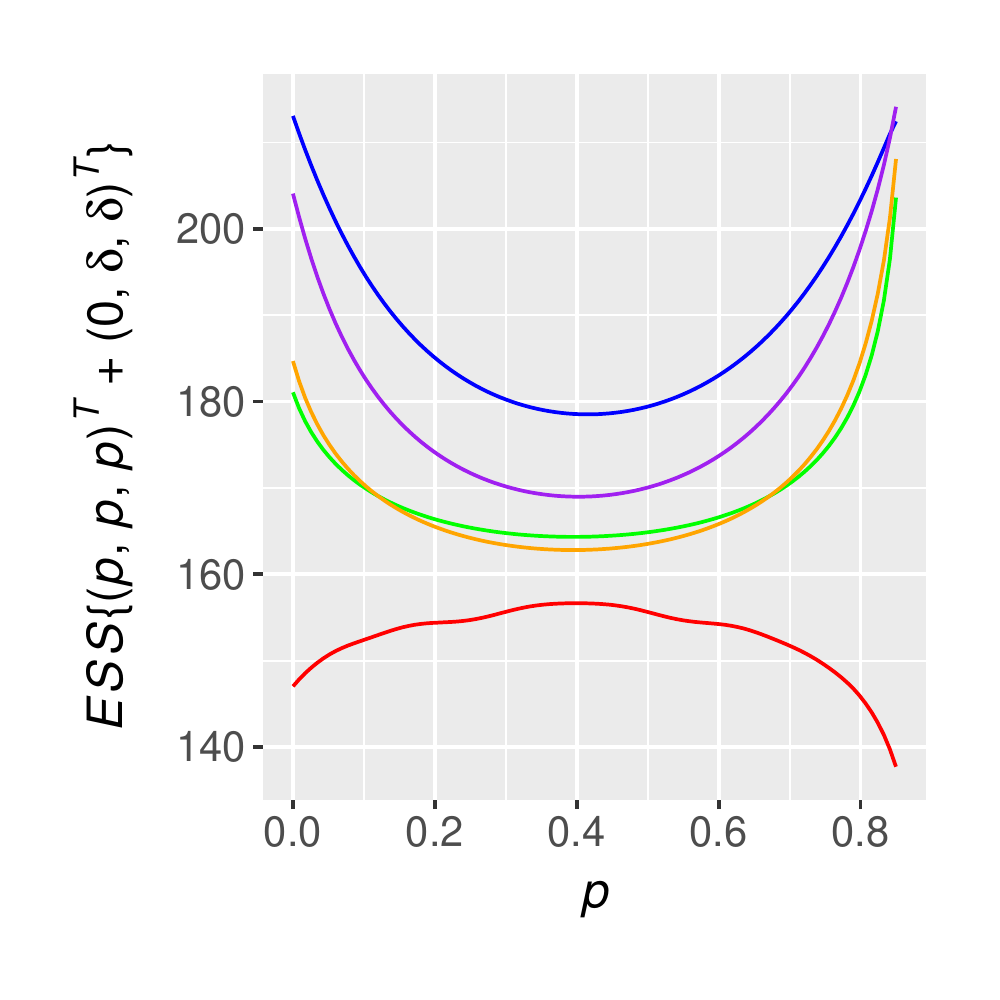}
}
\caption{The operating characteristics of the Fisher exact and exact binomial test designs given in Table 2 are presented. Specifically, in (a) and (c) the familywise error-rate (FWER) and expected sample size (ESS) when $\boldsymbol{p}=(p,\dots,p)^\top$ for $p\in[0,1]$ are shown respectively. In (b) and (d) the familywise power (FWP) and ESS when $\boldsymbol{p}=(p,\dots,p)^\top+\boldsymbol{\delta}$ for $p\in[0,0.85]$ are shown. The Fisher exact design is shown in red. The optimal exact binomial test designs with $\boldsymbol{w}=(1,0,0)$, $\boldsymbol{w}=(0,1,0)$, $\boldsymbol{w}=(1,1,0)$, and $\boldsymbol{w}=(0,1,1)$, are shown in blue, green, orange, and purple respectively.}
\end{figure}

\section{Discussion}\label{s:discussion}
	
In this paper, we proposed methodology for the design of a two-stage multi-arm randomised study with binary outcome variables, based upon Fisher's exact test. Through several examples, we were able to demonstrate that pre-specifying the stopping boundaries after stage one reduced potential trial efficiency substantially. Moreover, we observed that the Fisher exact design typically controlled the FWER more closely to the nominal level across possible values of the common success probability. In addition, this design typically retained a power advantage over its binomial exact counterparts, and would be expected to require a smaller sample size under the global alternative hypothesis. However, in some instances this did come at a small cost to the examined expected sample size under the global null hypothesis.

In general, the Fisher exact approach is advantageous over the exact binomial test since the common success probability under the global null hypothesis does not need to be searched over to weakly control the FWER. However, both methods considered here require a numerical search to control the type-II error rate to the desired level. 

The issues with controlling trial operating characteristics could of course be alleviated by the specification of point null and alternative hypotheses. However, in many design situations such an approach may not be wise. For example, randomised designs have been proposed for phase II clinical trials primarily to address issues caused by a lack of knowledge about the success probability in the control arm. It is worth noting that for each of the designs presented in Table 2, we utilised a stochastic search procedure to search for the maximal FWER across all $\boldsymbol{p}\in[0,1]^{K+1}$. In each case the maximum appeared to occur, as anticipated, when $\boldsymbol{p}=(p,\dots,p)^\top$ for some $p\in[0,1]$. This suggests strong control can be attained by controlling the maximal FWER over this $p\in[0,1]$. A formal proof of this fact remains to be presented however.

For both of the design procedures considered, it is extremely computationally intensive to determine the optimal design. Consequently, neither design can be declared preferential based on this factor. However, as discussed in Jung and Sargent (2014), the Fisher exact approach retains an advantage for scenarios in which a studies' sample size differs from its pre-specified value. This can easily be accounted for, by additionally conditioning on the realised sample sizes.

Throughout, we restricted ourselves to designs in which an equal number of patients were allocated to each of the experimental arms present in the study. This was necessary for the exact methods considered here, as unequal allocation would imply that different boundaries would be needed for each of the experimental arms. It is worth noting that the use of a normal approximation for the determination of optimal two-stage designs could more easily handle differing allocation to the experimental arms. Furthermore, this approach would readily extend to three-stage designs. Whereas, the exact methods discussed in this article may be computationally intractable with the addition of a third stage. However, a normal approximation approach would of course not be appropriate in the case of small sample sizes.

In conclusion, we have provided new methodology for the design of multi-arm studies with binary outcome variables. For the considered example, the operating characteristics of our design were found to often be preferable to an approach based on exact binomial tests. Thus, whilst there is no golden rule as to which technique should be preferred, our two-stage Fisher exact test may routinely be expected to provide more desirable efficiency gains.
	
\section*{Appendix}\label{appA}

The formal conduct of a trial utilising the exact binomial test approach is as follows
\begin{itemize}
	\item Set $\boldsymbol{\psi}=\boldsymbol{\omega}=\boldsymbol{0}$, and $j=1$.
	\item Conduct stage $j$ of the trial, allocating $(r_{Cj}-r_{Cj-1})n$ patients to the control arm, and $(r_{Ej}-r_{Ej-1})n$ patients to each arm $k\in\mathbb{N}_K$ with $\omega_k=0$. Following data accrual, compute the $T_{kj}$.
	\item For each $k\in\mathbb{N}_K$ with $\omega_k=0$
	\begin{itemize}
		\item If $T_{kj}\ge e_j$ reject $H_{0k}$, setting $\psi_k=1$ and $\omega_k=j$.
		\item If $T_{kj}\le f_j$ accept $H_{0k}$, setting $\omega_k=j$.
	\end{itemize}
	\item If $\sum_{k=1}^K\mathbb{I}\{\psi_k=1\}=0$ and $\sum_{k=1}^K\mathbb{I}\{\omega_k=0\}>0$, set $j=2$ and return to 2. Otherwise stop the experiment, and for each $k\in\mathbb{N}_K$ with $\omega_k=0$, set $\omega_k=j$.
\end{itemize}

With this, on trial termination we have that $(\boldsymbol{\Psi},\boldsymbol{\Omega})=(\boldsymbol{\psi},\boldsymbol{\omega})$. Then

\begin{align*}
\mathbb{P}(\boldsymbol{\omega},\boldsymbol{\psi}\mid \boldsymbol{p}) &= \sum_{x_{01}=0}^{r_{C1}n}\sum_{x_{11}=0}^{r_{E1}n}\dots\sum_{x_{K1}=0}^{r_{E1}n}\sum_{x_{02}=0}^{(r_{C2}-r_{C1})n}\sum_{x_{12}=0}^{(r_{E2}-r_{E1})n}\dots\\
& \qquad \sum_{x_{K2}=0}^{(r_{E2}-r_{E1})n} \left[ \prod_{k_1=1}^K \mathbb{I}\{T_{k_11}\in B_E(k_1,1,\boldsymbol{\psi},\boldsymbol{\omega})\} \right]\left[ \prod_{k_2=1}^K \mathbb{I}\{T_{k_22}\in B_E(k_2,2,\boldsymbol{\psi},\boldsymbol{\omega})\} \right]\\
& \qquad \qquad \left\{ b(x_{01}|r_{C1}n,p_0)\prod_{k_3=1}^Kb(x_{k_31}|r_{E1}n,p_{k_3}) \right\}\\
& \qquad \qquad \qquad \left[b\{x_{02}|(r_{C2}-r_{C1})n,p_0\} \prod_{k_4=1}^Kb\{x_{k_42}|(r_{E2}-r_{E1})n,p_{k_4}\} \right],
\end{align*}
where
\begin{equation*}
\text{B}_E(k,j,\boldsymbol{\psi},\boldsymbol{\omega}) = \begin{cases} (f_1,e_1) &:\ j<\omega_k, \\ (-\infty,\infty) &:\ j>\omega_k,\\ [e_j,\infty) &:\ j=\omega_k,\ \psi_k=1,\\ (-\infty,e_j-1] &:\ j=\omega_k=\max_l \omega_l,\ \psi_k=0,\ \sum_{l}\mathbb{I}(\psi_l=1)>0,\\ (-\infty,f_j] &:\ \text{otherwise}.\\  \end{cases}
\end{equation*}
	
Similarly, for the Fisher exact test design

\begin{align*}
\mathbb{P}(\boldsymbol{\omega},\boldsymbol{\psi}\mid \boldsymbol{p}) &= \sum_{x_{01}=0}^{r_{C1}n}\sum_{x_{11}=0}^{r_{E1}n}\dots\sum_{x_{K1}=0}^{r_{E1}n}\sum_{x_{02}=0}^{(r_{C2}-r_{C1})n}\sum_{x_{12}=0}^{(r_{E2}-r_{E1})n}\dots\\
& \qquad \sum_{x_{K2}=0}^{(r_{E2}-r_{E1})n} \left[ \prod_{k_1=1}^K \mathbb{I}\{T_{k_11}\in B_{F1}(k_1,z_1,\boldsymbol{\psi},\boldsymbol{\omega})\} \right]\left[ \prod_{k_2=1}^K \mathbb{I}\{T_{k_22}\in B_{F2}(k_2,z_1,z_2,\boldsymbol{\rho}_2,\boldsymbol{\psi},\boldsymbol{\omega})\} \right]\\
& \qquad \qquad \left\{ b(x_{01}|r_{C1}n,p_0)\prod_{k_3=1}^Kb(x_{k_31}|r_{E1}n,p_{k_3}) \right\}\\
& \qquad \qquad \qquad \left[b\{x_{02}|(r_{C2}-r_{C1})n,p_0\} \prod_{k_4=1}^Kb\{x_{k_42}|(r_{E2}-r_{E1})n,p_{k_4}\} \right],
\end{align*}
where
\begin{align*}
\text{B}_{F1}(k,z_1,\boldsymbol{\psi},\boldsymbol{\omega}) &= \begin{cases} (f_{1},e_{1z_1}) &:\ \omega_k=2, \\ [e_{1z_1},\infty) &:\ \omega_k=\psi_k=1,\\ (-\infty,f_{1}] &:\ \omega_k=1=\max_l\omega_l,\ \prod_l \mathbb{I}(\psi_l=0)=1,\\ (-\infty,e_{1z_1}-1] &:\ \omega_k=1=\max_l\omega_l,\ \psi_k=0,\ \prod_l \mathbb{I}(\psi_l=0)=0, \end{cases}\\
\text{B}_{F2}(k,z_1,z_2,\boldsymbol{\rho}_2,\boldsymbol{\psi},\boldsymbol{\omega}) &= \begin{cases} (-\infty,\infty) &:\ \omega_k=1, \\ [e_{2\boldsymbol{\rho}_2\cdot\boldsymbol{\rho}_2z_2z_1},\infty) &:\ \omega_k=2,\ \psi_k=1,\\ (-\infty,f_{2\boldsymbol{\rho}_2\cdot\boldsymbol{\rho}_2z_2z_1}] &:\ \omega_k=2,\ \psi_k=0.\end{cases}
\end{align*}
	
\section*{Acknowledgements}
This work was supported by the Medical Research Council [grant number MC\_UP\_1302/2 to M.J.G. and A.P.M.]; and the National Institute for Health Research Cambridge Biomedical Research Centre [MC\_UP\_1302/6 to J.M.S.W.].

\end{document}